\documentclass{aa}
\usepackage{graphicx}
\usepackage{txfonts}
\usepackage{natbib}
\bibpunct{(}{)}{;}{a}{}{,}
\begin{document}
   \title{The Metallicity of the Open Cluster Tombaugh 2.\thanks{Based on
          observations collected at ESO-VLT, Paranal Observatory, Chile, 
          Program numbers 076.D-0220(A)}}

   \author{S. Villanova\inst{1}
           \and
           S. Randich\inst{2}
           \and
           D. Geisler\inst{1}
           \and
           G. Carraro\inst{3,4}
           \and
           E. Costa\inst{5}
           }

   \institute{Universidad de Concepci\'on, Departamento de Astronomia, Casilla
              160-C, Concepci\'on, Chile\\
              \email{svillanova@astro-udec.cl,dgeisler@astro-udec.cl}
              \and
              INAF/Osservatorio Astrofisico di Arcetri, Largo E. Fermi 5,
              50125 Firenze, Italy\\
              \email{randich@arcetri.astro.it}
              \and
              ESO, Alonso de Cordova 3107, Vitacura, Santiago, Chile\\
              \email{gcarraro@eso.org}
              \and
              Dipartimento di Astronomia, Universit\'a di Padova,
              Vicolo Osservatorio 3, I-35122, Padova, Italy
              \and
               Departamento de Astronom\'ia, Universidad de Chile,
               Casilla 36-D, Santiago, Chile\\
              \email{costa@das.uchile.cl}
              }

   \date{Received -----------; accepted -----------}

% \abstract{}{}{}{}{} 
% 5 {} token are mandatory
 
  \abstract
   {Open clusters are excellent tracers of the structure, kinematics, and 
chemical evolution of the disc and a wealth of information can be derived from
the spectra of their constituent stars.
}
   {We investigate the nature of the chemical composition of the outer disc open
cluster Tombaugh 2. This has been suggested to be a member of the GASS/Mon
substructure, and a recent study by Frinchaboy et al. (2008)
suggested that this was a unique open cluster
in possessing an intrinsic metal abundance dispersion. We aim to investigate 
such claims.
}
   {High resolution VLT+GIRAFFE spectra in the optical are obtained and analyzed
for a number of stars in the Tombaugh 2 field, together with independent
UBVI$_{\rm C}$ photometry. Radial velocities and position in the color-magnitude 
diagram are used to assess cluster membership.  The spectra, together 
with input atmospheric parameters and model atmospheres, are used
to determine detailed chemical abundances for a variety of elements in 13 
members having good spectra.
}
   {We find the mean metallicity to be [Fe/H]$=-0.31 \pm 0.02$ with no evidence
for an intrinsic abundance dispersion, in contrary to the recent results of 
Frinchaboy et al. (2008). We find Ca and Ba to be slightly enhanced while Ni and Sc are
solar. The r-process element Eu was found to be enhanced, giving an average
[Eu/Ba]=+0.17. The Li abundance decreases with T$_{\rm eff}$ on the upper giant branch and
maintains a low level for red clump stars. The mean metallicity we derive is 
in good agreement with that expected from the radial abundance gradient in the
disc for a cluster at its Galactocentric distance. 
}
   {Tombaugh 2 is found to have abundances  as expected from its Galactocentric
distance and no evidence for any intrinsic metallicity dispersion. The 
surprising result found by Frinchaboy et al. (2008), that is the presence of 2 distinct abundance
groups within the cluster, implying either a completely unique open cluster
with an intrinsic metallicity spread, or a very unlikely superposition of a 
cold stellar stream and a very distant open cluster, is not supported by our
new result.
}


   \maketitle
%
%________________________________________________________________

\section{Introduction}

\begin{figure*}
\centering
\includegraphics[width=\columnwidth]{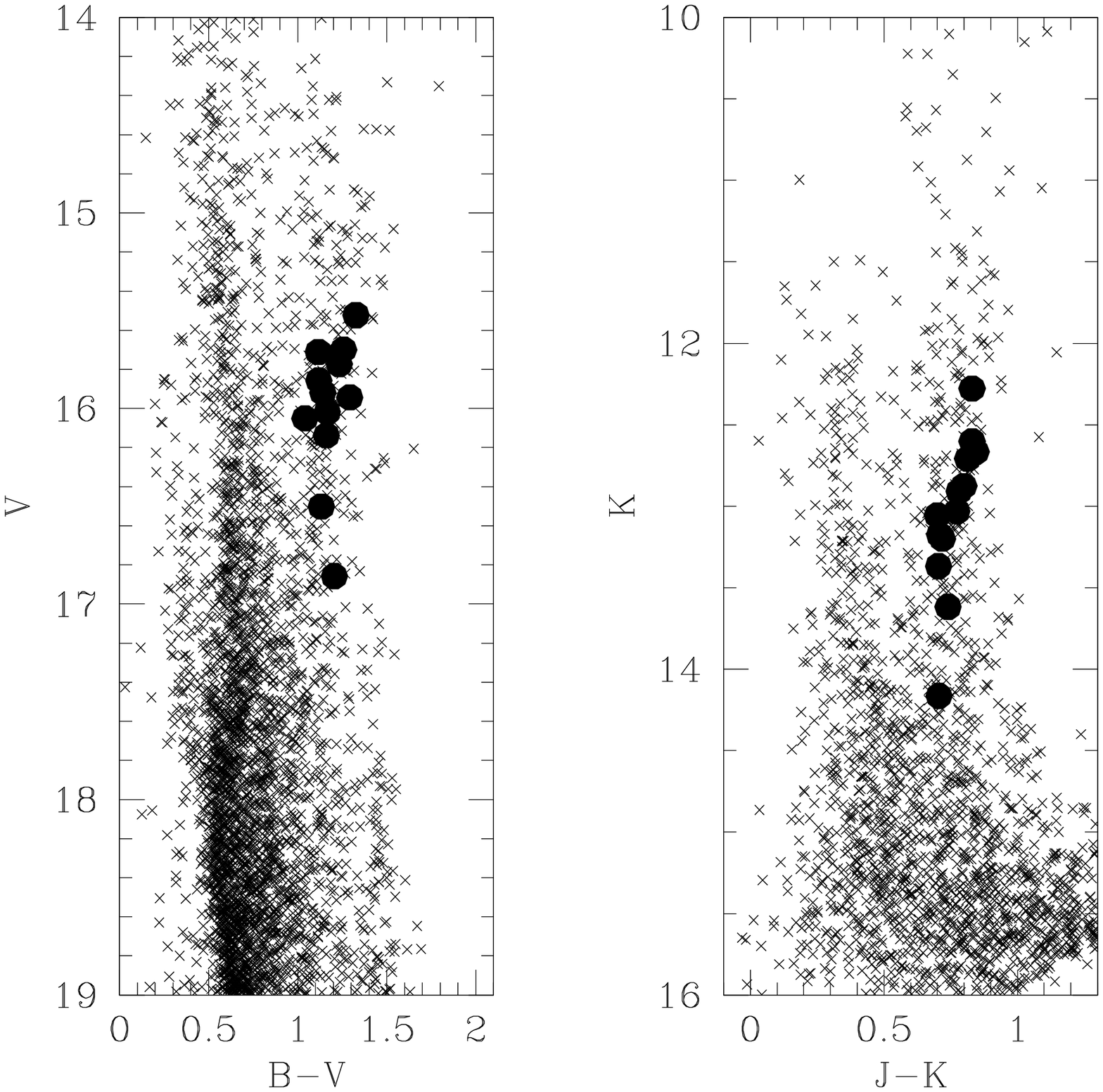}
\includegraphics[width=\columnwidth]{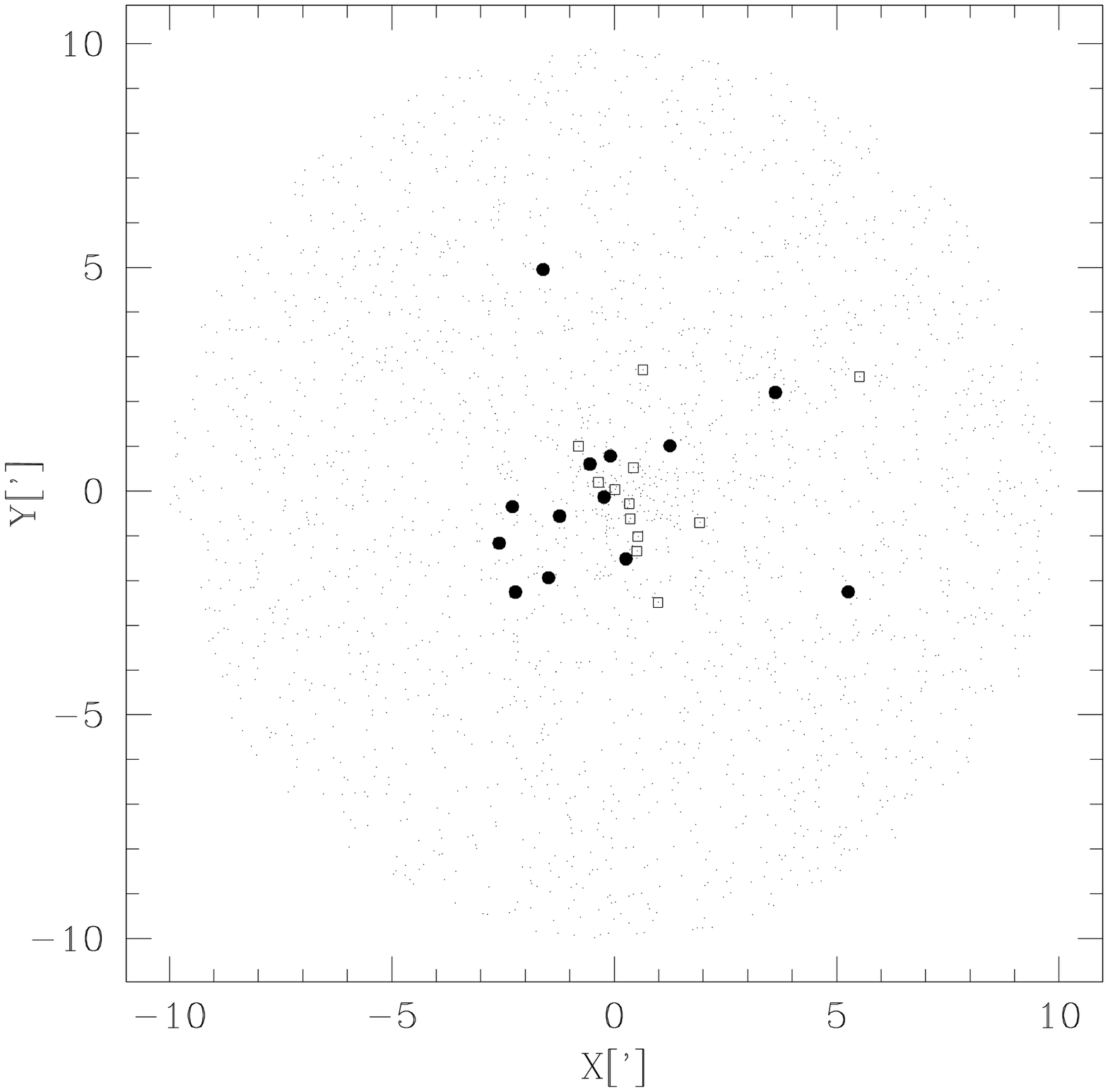}
\caption{Left panel: The CMD of Tombaugh 2 in the optical and infrared.
Members stars studied in this paper are indicated as filled circles.
Right panel: Distribution on the sky of our targets (filled points) and F08
objects (empty squares).}
\label{f1}
\end{figure*}

One of the many regions in our Galaxy for which we lack detailed knowledge is
the outer disk. How and when was it formed? Via an outside-in or inside-out
process \citep{chi01}? What is the nature of the
metallicity gradient? Does it show a constant slope with Galactocentric distance
or is there a leveling out in the outer disk \citep{twa97,car04,car07,magr09}
What if any is the role of mergers? \\
A particularly intriguing development in recent years has been the suggestion 
that there is a merger remnant lying near the Galactic plane. The structure
was first identified as the Monoceros stream \citep[Mon;][]{new02,iba03,yan03},
also known as the Galactic anticenter stellar structure  \citep[GASS;][]{roc03,cra03}
GASS/Mon was discovered as an overdensity of stars near the Galactic plane
that seems to wrap around the outer parts of the Galactic disc and is frequently
explained as tidal debris from the disruption of a dwarf galaxy on a low inclination orbit
\citep[e.g.,][]{cra03,yan03,pen05}.\\
Further arguments have been made that
the reputed dwarf galaxy in Canis Major \citep[CMa;][]{bel04}
is the progenitor of the Mon/GASS structure \citep{martin04}.
However, the nature and/or reality of the proposed Mon ``tidal debris stream'' and the CMa
overdensity have been called into question and are currently a matter of great debate.
Momany \citeyearpar{mom04,mom06} have argued that much of the observed stellar
overdensity associated with Mon --- and particularly all of that associated with
CMa (at $l \sim 240^{\circ}$) ---
is due to the warping and flaring of the Galactic disc, and that no ``extra-Galactic'' component is
needed to account for the apparent overdensities in the third quadrant. 
The presence of `blue plume'' stars in this part of the sky
has been used to argue further for the presence of a dwarf galaxy nucleus in CMa
\citep{bel04,martinez05,din05,but07}.
However, these young stars have also been more prosaically attributed to the 
presence of previously unknown features of spiral arm
structure \citep{car05,moi06}.

Open clusters have traditionally played a critical
role in the study of the 
structure, kinematics, formation and chemical evolution of the disk. For this
reason alone, the old outer Galactic disk open cluster Tombaugh 2 (To2;
at Galactic coordinates [$l,b$] = [232.8,$-$6.9]$^{\circ}$), 
located at a Galactocentric distance of $\sim$15 kpc and 700 pc below
the Galactic plane, is worthy of study
and was indeed the subject of several previous metallicity investigations.
\citet{bro96} obtained high resolution spectra of 3 stars, finding 
[Fe/H]$=-0.4\pm 0.25$, while \citet{frie02} derived [Fe/H]$-0.44 \pm 0.09$ from
much lower resolution spectra of 12 members. Both of these studies were 
conducted with the CTIO 4m telescope.
Subsequently, \citet{frin04} suggested that To2
may possibly be associated with
Mon/GASS, as well as with CMa \citep{bel04}. Note that its Galactic
longitude is very close to that of the CMa overdensity.
This provided a  strong additional 
incentive  for a followup study of
 its chemistry and kinematics in greater detail with a 
larger telescope, as even the giants are faint due to its distance of over 13kpc.
\citet[][ - hereafter F08]{frin08} used this motivation to obtain
VLT/FLAMES spectra (both with UVES at high resolution and with GIRAFFE at 
somewhat lower resolution). They were able to derive velocities and detailed
abundances of a number of elements for 18 To2 cluster members. 
Their most surprising result was an apparent large spread in metallicity
: $\Delta$[Fe/H] $> 0.2$. They were unable to account for this spread given
their observational errors and presented a number of possible scenarios,
including the likelihood that To2 possessed an intrinsic metallicity spread.
They argued for the possible
presence of 2 populations in the cluster, distinguished by their 
different mean chemical characteristics --- with a metal-rich, Ti-normal group
and a metal-poor, Ti-enhanced group, 
namely ($\langle$[Fe/H]$\rangle$, $\langle$[Ti/Fe]$\rangle$) $\sim$
($-$0.06, $+$0.02) and ($-$0.28, $+$0.36). The more metal-poor group appeared
more centrally concentrated, and they suggested that this group represented
the true To2 clusters stars and the
metal-rich population was   an overlapping,
and kinematically associated, but ``cold''  stellar stream. 

If true, this would be the first such metallicity
spread uncovered in an open cluster, and/or the first evidence of multiple 
populations in a cluster of this age (about 2 Gyr) in the Galaxy,
making To2 an extremely intriguing object. 
Although multiple populations within globular clusters are now known
to exist, these clusters are all much more massive than an open cluster such
as To2, as expected if this phenomenon is due to a cluster being massive 
enough to retain ejecta from a first generation of stars in order to make a 
second, chemically enriched generation.      To date, the F08 study is the only 
one which suggests multiple populations in an open cluster.
However, as F08 pointed out, further 
observations are desperately required to corroborate their surprising results
and make a more definitive investigation.

Given its above history, a new metallicity study of To2 is imperative.
These points motivated the present study, wherein we wished to obtain 
independent high-resolution spectra with a large telescope of a number of To2
stars in order to investigate the reality of the putative metallicity dispersion
and the nature of its chemical composition.
In 2, we describe our observations and reduction procedures. In 3, we discuss
the determination of radial velocities and membership for our observed sample.
In 4, we present the details of our abundance analysis, including the derivation
of the input atmospheric parameters and of the internal errors. In 5, our basic abundance results are
presented, and in 6 we compare these to previous investigations, especially
that of F08 in light of the above. Finally, we summarize and emphasize the
importance of our main results in 7.

\begin{table*}
\caption{Target stars and parameters}
\label{t1}
\centering
\begin{tabular}{llccccccccccccc}
\hline
\hline
ID1 & ID2 & $\alpha$($^0$) & $\delta$($^0$) & V & U-B & B-V & V-I$_{\rm C}$ & J$_{\rm 2MASS}$ &\
H$_{\rm 2MASS}$ & K$_{\rm 2MASS}$ & T$_{\rm eff}$ & log g & v$_{\rm t}$ & RV$_{\rm H}$\\
\hline                      
1512 & 2138 & \tiny{105.744536} & \tiny{-20.848913} & 16.857 & 1.122 & 1.206 &  1.294 & 14.870 & 14.278 & 14.164 & 5210 & 3.29 & 1.10 & 117.6\\
1672 & 1100 & \tiny{105.724763} & \tiny{-20.836042} & 15.726 & 0.818 & 1.231 &  1.319 & 13.511 & 12.850 & 12.666 & 4820 & 2.63 & 1.26 & 119.1\\
1827 &  837 & \tiny{105.748951} & \tiny{-20.825960} & 15.773 & 0.877 & 1.233 &  1.347 & 13.521 & 12.865 & 12.708 & 4900 & 2.70 & 1.24 & 120.9\\
1886 & 1532 & \tiny{105.730038} & \tiny{-20.822460} & 15.711 & 0.638 & 1.116 &  1.251 & 13.675 & 13.024 & 12.876 & 5010 & 2.72 & 1.24 & 120.0\\
2184 &  485 & \tiny{105.793143} & \tiny{-20.799812} & 15.857 & 0.702 & 1.118 &  1.262 & 13.755 & 13.214 & 13.055 & 4980 & 2.77 & 1.23 & 119.8\\
 236 & 2895 & \tiny{105.864625} & \tiny{-20.854195} & 15.919 & 0.669 & 1.141 &  1.281 & 13.804 & 13.234 & 13.031 & 5050 & 2.83 & 1.21 & 121.0\\
 238 & 1802 & \tiny{105.731226} & \tiny{-20.854284} & 15.699 & 0.889 & 1.258 &  1.341 & 13.431 & 12.771 & 12.601 & 4780 & 2.60 & 1.27 & 120.5\\
2846 & 2430 & \tiny{105.742321} & \tiny{-20.734119} & 16.017 & 0.547 & 1.166 &  1.187 & 14.072 & 13.526 & 13.368 & 5160 & 2.92 & 1.19 & 122.9\\
 299 & 2902 & \tiny{105.775540} & \tiny{-20.841908} & 15.523 & 1.002 & 1.327 &  1.426 & 13.107 & 12.397 & 12.277 & 4780 & 2.53 & 1.29 & 121.8\\
3574 & 2074 & \tiny{105.766791} & \tiny{-20.818871} & 16.138 & 0.678 & 1.161 &  1.321 & 13.919 & 13.311 & 13.200 & 5150 & 2.96 & 1.18 & 121.7\\
3763 & 2894 & \tiny{105.761083} & \tiny{-20.806566} & 16.051 & 0.783 & 1.041 &  1.288 & 13.883 & 13.367 & 13.175 & 5000 & 2.86 & 1.20 & 120.1\\
3836 &  494 & \tiny{105.769301} & \tiny{-20.803593} & 16.501 & 0.704 & 1.133 &  1.382 & 14.357 & 13.750 & 13.617 & 5110 & 3.09 & 1.15 & 120.9\\
 591 & 2442 & \tiny{105.835422} & \tiny{-20.779896} & 15.944 & 0.706 & 1.293 &  1.355 & 13.690 & 13.018 & 12.909 & 5020 & 2.82 & 1.21 & 125.1\\
\hline     
\end{tabular}
\end{table*}

\section{Observations and data reduction}
Our data-set consists of high resolution spectra collected
with FLAMES-GIRAFFE/VLT@UT2 \citep{pas02} in Service
mode from March 6 to March 25 2006,
within a project devoted to measure radial velocities, 
membership, and chemistry in a large sample of open clusters \citep{rand05}. 
The GIRAFFE spectrograph was used in the HR15N setting, providing a resolution 
R$\sim 17,000$ and covering
a spectral range of $\sim$320 \AA \ with the central wavelength at 6650 \AA.
Typical seeing during the observations was in the range 0.8-1.2 arcsec.

The cluster was observed with
two different configurations (A and B), centered at the same position
(RA(2000)=07h~03m~01.95s, DEC=$-$20d~49m~50.2s). We obtained four and 
three 45~min long exposures for configurations A and B, respectively.
Medusa fibers were allocated to 93 and 120 stars in the two configurations,
with 78 stars in common. Hence, we obtained in total
spectra of 135 cluster candidates.
These cover the magnitude range $14 \leq \rm V \leq 18.7$
and are located in different regions of the cluster color-magnitude
diagram (CMD): namely, the
turn-off and subgiant branch, the red giant branch (RGB), the red clump (RC), 
and the blue plume.
In this paper we focus on RGB and RC stars for a total of 37 objects. 15 of them
turned out to be members (see below for the membership determination) but only
13 had spectra with sufficient quality in order to measure chemical abundances
and are listed in Table~\ref{t1}. Their position in the cluster CMDs is
shown in Fig.~\ref{f1}.
Targets were originally selected from \citet{phe94} photometry.
However in this paper we use our new optical photometry (see following sub-section)
and 2MASS (JHK --\citep{skr06}).\\
Data were reduced using GIRAFFE pipelines (Ballester et al.\ 2000),
including bias subtraction, flat-field correction, and wavelength
calibration. Sky subtraction was performed plate by plate using the median sky as obtained
from the fibers pointed on empty regions of the field.
Radial velocities were
derived from each single spectrum (see below). Spectra of stars that were
not found to be RV~variables were then co-added.
The typical signal to noise ratio per pixel is ${\rm S/N\sim 60-80}$.

\begin{figure*}
\centering
\includegraphics[width=9cm]{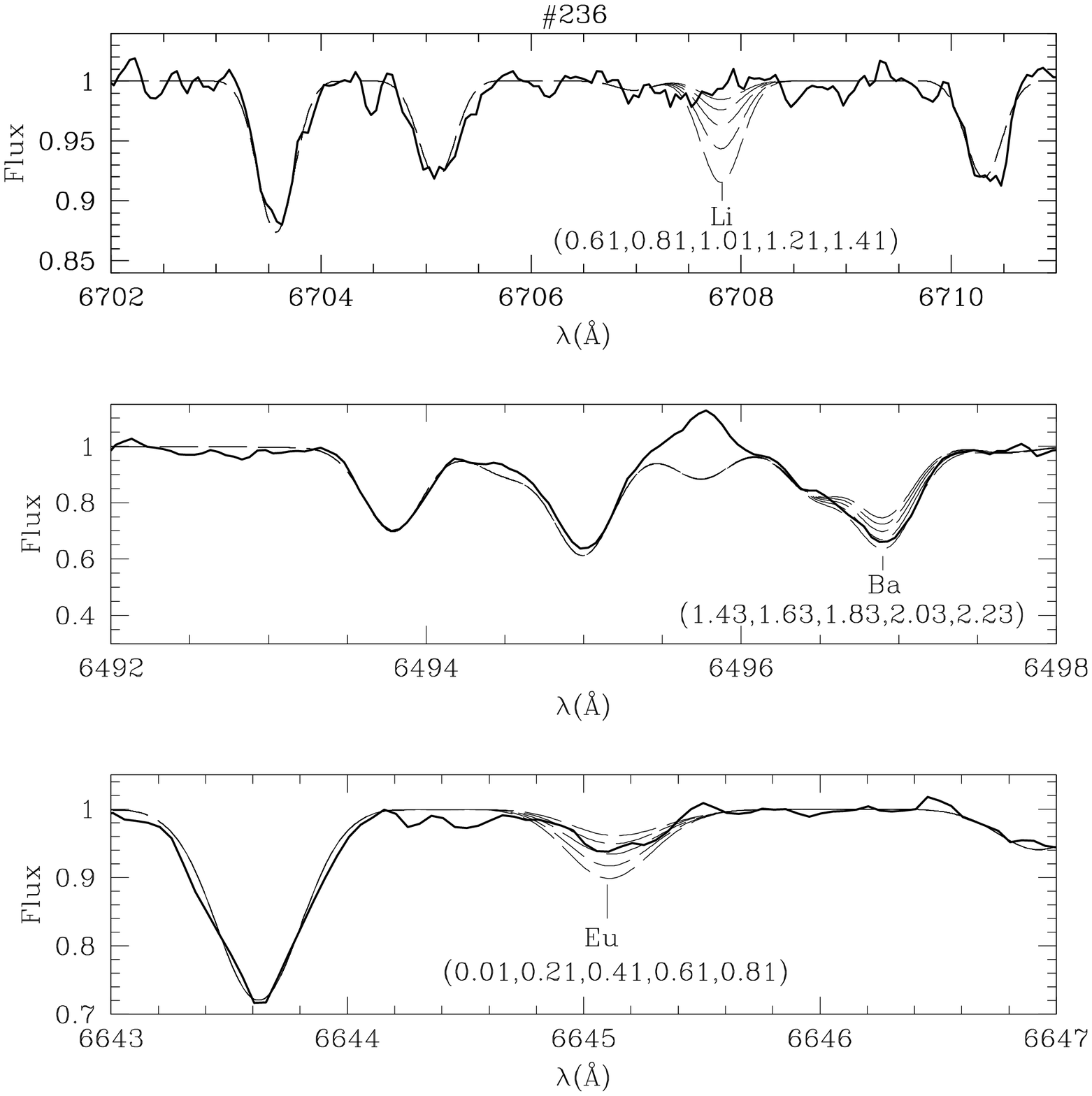}
\includegraphics[width=9cm]{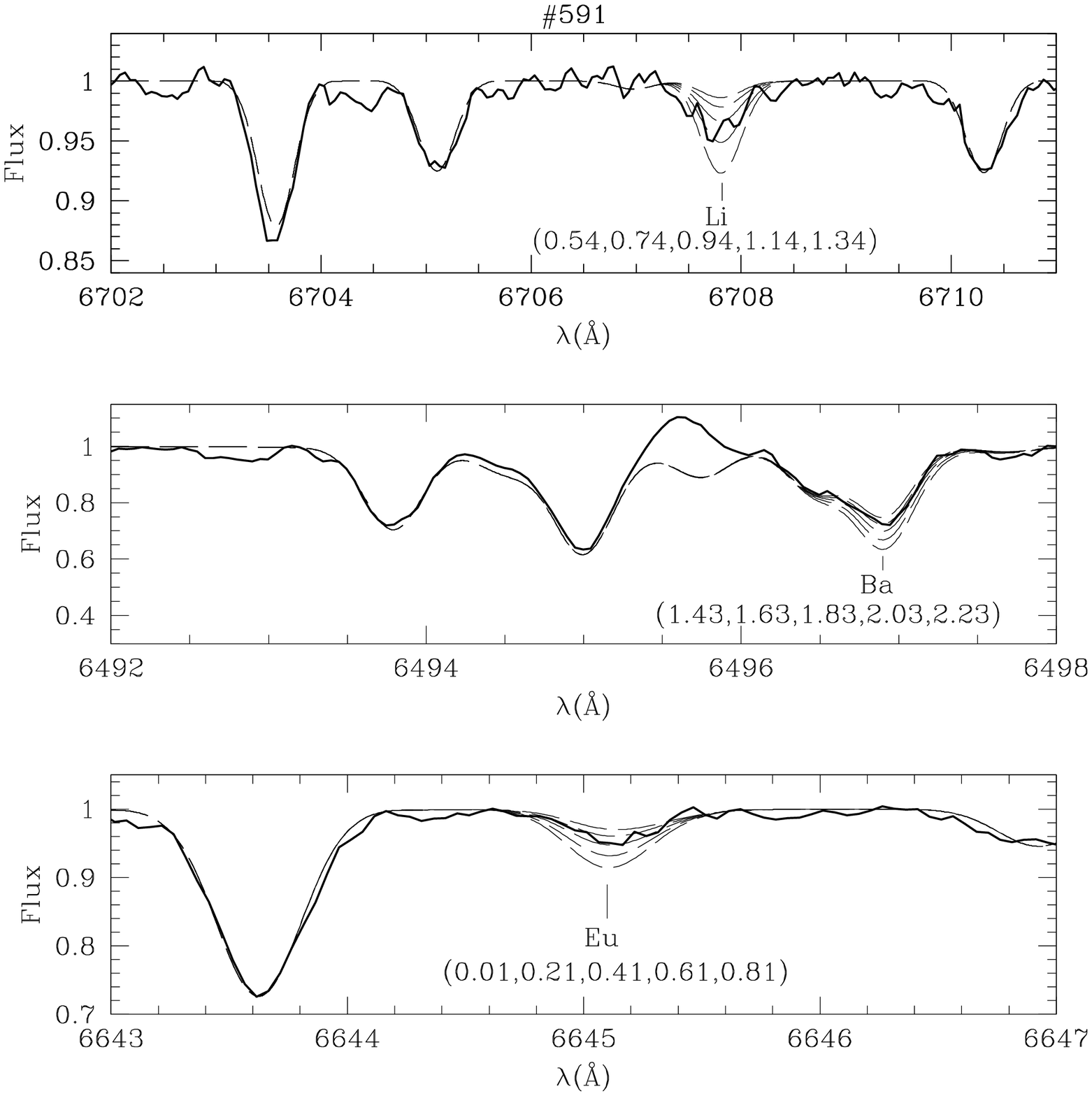}
\caption{Example of the spectral synthesis method applied to stars \#236 and
  \#591. Observed spectra are plotted with continuous lines, while synthetic
  with dashed lines. Abundances ($\rm {log\epsilon(X)=log(N_{X}/N_{H})+12)}$)
  used in the synthesis are indicated.}
\label{f2}
\end{figure*}

\subsection{Photometric material}
We complemented the spectroscopic observations with 
new UBVI$_{\rm C}$ observations secured with the Cerro 
Tololo Inter-American Observatory 1.0m telescope,
operated by the SMARTS consortium, as a part of a large photometric program to study
stellar fields in the third Galactic Quadrant.
Tombaugh~2 was observed on Christmas night of 2008 in photometric conditions.
Full details of the observations, data reduction and photometric
calibration can be found in \citep{car09} and references therein. 
Here we would just like to point out
the large field (20$\arcmin$ on a side) and good scale (0.297$\arcsec$/pixel) provided
by our set-up, which allows for an optimum study of both the central cluster and the surrounding
field. For the purpose of this study, we selected stars within 3 arcmin from the nominal
cluster center, and cross-correlate our data with the 2MASS catalog.
Typical deep exposures times 
in U, B, V and I$_{\rm C}$ were 2000, 1500, 1200, and 1200 sec, respectively.\\

\section{Radial velocities and membership}
In the present work, radial velocities, coupled with the position of the star
along the principal giant sequences in the cluster color-magnitude diagram,
 were used as the membership criterion
since cluster stars all have similar motion with respect to the observer.
The radial velocities of the
stars were measured using the IRAF FXCOR task, which cross-correlates
the object spectrum with a template.  As a template, we used a
synthetic spectrum obtained through the spectral synthesis code
SPECTRUM\footnote{see {\sf http://www.phys.appstate.edu/spectrum/spectrum.html} for
more details}, using a Kurucz model atmosphere \citep{kur92} with roughly the mean
atmospheric parameters of our stars $\rm {T_{eff}=5000}$ K, $\rm {log g=2.9}$,
$\rm {v_{t}=1.2}$ km/s, $\rm {[Fe/H]=-0.30}$. At the end, each radial velocity was
corrected to the heliocentric system. We calculated a first approximation
mean velocity and the r.m.s ($\sigma$) of the velocity distribution.
Stars showing $\rm {v_{r}}$ more than
$3\sigma$ from the mean value were considered probable field objects and
rejected, leaving us with 15 objects as probable members. Only
13 of them had spectra with sufficient quality in order to measure chemical
abundances and their positions in the CMD and on the sky are shown in Fig.~\ref{f1}. 
Radial velocities of these 13 targets are reported in Tab.~\ref{t1}, while
coordinates, V magnitudes and velocities for non-members and the two member
with bad spectra are reported in Tab.~\ref{t5}. We found for the cluster
a mean heliocentric radial velocity and an observed dispersion of:

\begin{center}
${\rm <v_r>=120.9\pm0.4,\ \sigma_{v_r}=1.7\pm0.3\ km/s}$
\end{center}

Both of these values are in excellent agreement with F08.

Our analysis shows that only 15 out of the 37 sample stars are cluster members,
thus yielding a contamination of $\sim 60$~\%.

\section{Abundance analysis}

\subsection{The linelist}

The linelist for the elements we measured (with the exception of Li, Ba, and Eu)
were obtained from the VALD
database \citep{kup99} and calibrated using the Solar-inverse technique.
For this purpose we used the high resolution, high S/N Solar spectrum
obtained at NOAO ($National~Optical~Astronomy~Observatory$, Kurucz et
al.\ 1984). The EWs for the reference Solar spectrum were obtained in the same
way as the observed spectra (see next subsection), with the exception of the strongest lines, where
a Voigt profile integration was used. Lines affected by blends were rejected
from the final line-list.
Abundances were determined using MOOG\footnote{see {\sf 
http://verdi.as.utexas.edu/moog.html} for more details},
coupled with a solar model atmosphere interpolated from the \citet{kur92} grid
using the canonical atmospheric parameters for the Sun: $\rm {T_{eff}=5777}$ K,
$\rm {log g}=4.44$, $\rm {v_{t}=0.80}$ km/s and $\rm {[Fe/H]=0.00}$.
In the calibration procedure, we adjusted the value of the line strength
log(gf) of each spectral line so that  the abundances obtained from
all the lines of the same element yield the same value.
Note that this procedure is identical to that used in F08.\\
For Ba, known to be affected by hyperfine structure and isotopic effects, we
applied the spectrum synthesis method instead. For the Sun we considered Ba lines at
4554, 5853, 6141, and 6496 \AA. Hyperfine components of those lines and
isotopic composition were taken from McWilliam (1998). 
For each line we determined a Ba abundance, and then we adjusted the 
log(gf) value of each one in order to yield  the same abundance obtained from
all the lines. For the present data we used only the Ba
feature at 6496 \AA.\\
The Eu line we used is located at 6645 \AA\ and it is blended with a Si
feature. We verified that at the temperature of our stars the contribution of
the Si line is negligible. Anyway we obtain Eu abundance by spectral synthesis method.
For Si abundance, lacking a direct determination, we assumed the
$\alpha$-enhancement as obtained by Ca. The Eu line is so weak that it does not
require a hyperfine structure treatment.\\
The Li line at 6707\ \AA\ is an unresolved doublet, so we performed a spectrum
synthesis analysis also for this element.\\
For Eu and Li we took parameters for the 6707 \AA\ line from VALD and 
NIST \footnote{NIST database can be found at http://physics.nist.gov/PhysRefData/ASD/lines\_form.html}
databases, and the log(gf) values simply averaged.\\
Chemical abundances obtained for the Sun are reported in Tab.~\ref{t2} and
compared with \citet{gre98}. For Li we report the meteoritic value. 
We use our abundaces instead of the those from \citet{gre98} or \citet{asp05}
because they were obtained in an homogenous way with respect to our target stars.

\subsection{Continuum, atmospheric parameters and chemical abundance determination}

The chemical abundances for Li, Ba, and Eu were obtain by comparing
observed spectra with synthetic ones (see Fig.~\ref{f2} for an example of the
spectral synthesis applied to stars \#236 and \#591), so in this case 
the continuum detrermination is a by-product of this procedure.
For the remaining elements abundances were obtained from the equivalent widths
(EWs) of the spectral lines. 
In this second case continuum determination is a bit more complicated. Our
spectra are centered on the H$_{\alpha}$ region,
so they are relatively free from spectral lines, especially compared to the data
of F08, whose spectral range was much bluer and contains many more lines.
This fact allowed us to proceed in the following way. First, 
for each line, we selected a region of 20 \AA \ centered on the line itself
(this value is a good compromise between having enough points, i. e.  good statistics, and 
avoiding an excessively large spectral region over which the spectrum 
could be substantially curved).
Then we built the histogram of the distribution of the flux where the peak is a
rough estimation of the continuum. We refined this determination by fitting a
parabolic curve to the peak and using the vertex as our continuum estimation. 
Finally, the continuum determination was revised by eye and corrected by hand
if a clear discrepancy with the spectrum was found.
Then, using the continuum value previously obtained, we fit a Gaussian curve
to each spectral line and obtained the EW from integration.
We rejected lines if affected by bad continuum determination, by non-Gaussian
shape, if their central wavelength did not agree with that expected from 
our line-list, if the lines were too broad or too narrow with respect to the
mean FWHM, or if it was affected by blending with other spectral features.
We verified that the Gaussian shape was a good approximation for our (mostly
weak) spectral lines, so generally no Lorentzian correction was applied.
The used lines and the measured EWs are reported in Tab.~\ref{t6}.

Initial estimates of the atmospheric parameters were derived from 
BVI$_{\rm C}$JHK photometry. Effective temperatures (T$_{\rm eff}$) for each 
star were derived from T$_{\rm eff}$-[Fe/H]-color relations
\citep{alo99,dib98,ram05}.
Colors were de-reddened assuming a reddening value of 0.3
mag \citep{frin08}.
Then T$_{\rm eff}$ was adjusted in order that abundances derived for
individual FeI lines show no trend with excitation potential.

\begin{figure*}
\centering
\includegraphics[width=12cm]{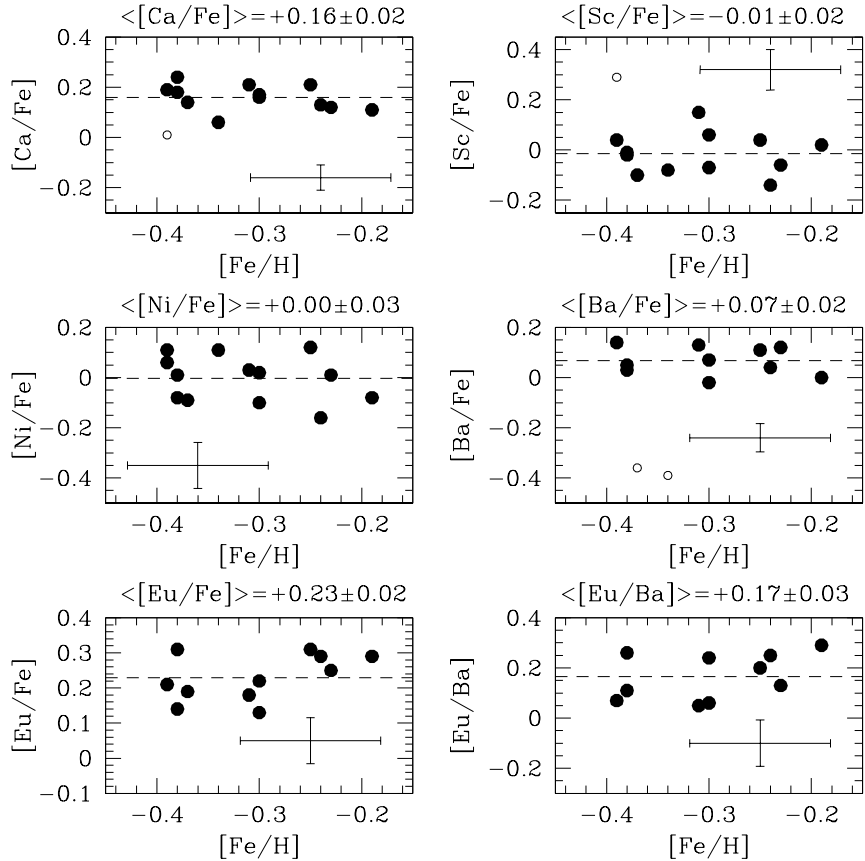}
\caption{Abundance ratios obtained for our stars. Typical error bars and mean
  values (dashed lines) are indicated. Open circles are values rejected during
  the sigma clipping rejection procedure (see Sec. 5).}
\label{f3}
\end{figure*}

Surface gravities log g were obtained from the canonical equation:

\begin{center}
${\rm log(g/g_{\odot}) = log(M/M_{\odot}) + 4\cdot
  log(T_{eff}/T_{\odot}) - log(L/L_{\odot}) }$
\end{center}

For M/M$_{\odot}$ we adopted 1.45 M$_{\odot}$ based on our isochrone fitting.

The luminosity ${\rm L/L_{\odot}}$ was obtained from the absolute
magnitude M$_{\rm V}$ using the measured V magnitude, assuming the bolometric
correction ($\rm {BC}$) from \citet{alo99}, and an apparent distance modulus 
(m-M)$_{\rm V}$ obtained in the following way.\\
For many of the spectra we could measure the FeII line at 6456 \AA. While just one FeII line is not
enough to determine a reliable gravity for a single star, it can be used 
to determine the apparent distance modulus (which is the same for all the
stars), simply varying it until the mean FeI and FeII abundances of the cluster (calculated
assuming the gravity obtained from the previous formula) are the same.
All other quantities in the gravity equation (T$_{\rm eff}$,
M/M$_{\odot}$, BC, V) are known.\\
We obtained:

\begin{center}
${\rm (m-M)_{V}=15.01\pm0.12}$
\end{center}

\begin{table}
\caption{Our measured Solar abundances ($\rm
  {log\epsilon(X)=log(N_{X}/N_{H})+12)}$ compared with \citet{gre98}. 
  For Li we use the meteoritic value.} 
\label{t2}
\centering
\begin{tabular}{lcc}
\hline
\hline
Element &  log$\epsilon$(X) & GS98\\
\hline
LiI &  -   &  3.31 \\
CaI & 6.39 &  6.36 \\
ScI & 3.12 &  3.17 \\
FeI & 7.50 &  7.50 \\
NiI & 6.28 &  6.25 \\
BaI & 2.34 &  2.13 \\
Eu  & 0.51 &  0.52 \\
\hline
\end{tabular}
\end{table}

We point out that this value was optimized in order to give [FeI/H] = [FeII/H],
so it can differ from the real apparent distance modulus because the other
variables in the gravity equation (mainly the mass) can be affected by
systematic errors. This can explain the difference with F08, where the authors
find (m-M)$_0\sim$14.5, while we have (m-M)$_0\sim$14.30 assuming E(B-V)=0.23
(see section 5). However the two values are not incompatible within the
errors.\\
Finally, microturbulence velocity ($\rm {v_{t}}$) was obtained from the
relation:

\begin{center}
${\rm v_{t}\ (km/s) = -0.254\cdot log g+1.930}$
\end{center}

which was obtained  from \citet{marino08}.\\
Adopted atmospheric parameters for each star are reported in Tab.~\ref{t1}.
In this Table column 1 and 2 give the ID of the star according to
\citet[][ID1]{phe94} and our photometry (ID2). Columns 3 and 4 the
coordinates, columns 5-11 the B,V,I,J,H,K magnitudes, column 12 the temperature
(K), column 13 the gravity, column 14 the microturbulence velocity (km/s), and column 15 the
heliocentric radial velocity (km/s).\\
The Local Thermodynamic Equilibrium (LTE) program MOOG was used to
determine the abundances, coupled with a model atmosphere
interpolated from the Kurucz models for the parameters obtained as described
before. Results are reported in table~\ref{t3}.

\subsection{Internal errors associated with the calculation of the chemical abundances}

The abundances of every element is affected by measurement errors.
In this section our goal is to estimate the total amount of internal error
($\sigma_{\rm {tot}}$) affecting our data. Clearly, this requires an accurate
analysis of all the internal sources of errors.
External errors can be estimated by comparison with other works, as done in
section 6.\\
It must be noted that two sources of error mainly contribute
to $\sigma_{\rm {tot}}$. They are:
\begin{itemize}
\item the errors $\sigma_{\rm {EW}}$ due to the uncertainties in the
  EWs measures or in the comparison of observed 
  spectra with synthetic ones.
\item the uncertainty $\sigma_{\rm {atm}}$ introduced by errors in the 
  atmospheric parameters adopted to compute the chemical abundances.
\end{itemize}

$\sigma_{\rm {EW}}$ is given by MOOG for each element and each star in the
case of Ca, Fe, and Ni. For Li, Ba, and Eu, whose abundances were obtained by
the spectral synthesis method. $\sigma_{\rm {EW}}$ was obtained using the
$\chi^2$ method, as described in \citet{vil09}. 
In Tab.~\ref{t4} we report in the second column the average $\sigma_{\rm {EW}}$
for each element. For Sc we were able to measure only one line.
For this reason its $\sigma_{\rm {EW}}$ has been obtained as the mean of
$\sigma_{\rm {EW}}$ of Ni multiplied by $\sqrt 2$. 

Errors in temperature were determined as in \citet{marino08}. The mean
error $\Delta$T$_{\rm eff}$ turned out to be $\sim$50 K.
Uncertainty in surface gravity has been obtained by the canonical formula using the
propagation of errors. The variables used in this formula that are affected
by random errors are T$_{\rm eff}$ and the V magnitude. For temperature we
used the error previously obtained, while for V we assumed a mean error of 0.1 mag,
which is the typical random error for stars at that magnitude. Other error
sources (distance modulus, reddening, bolometric correction, mass) affect
gravity in a systematic way, so are not important to our analysis.
The mean gravity error turned out to be 0.06 dex. This implies a mean error
in the microturbulence of 0.02 km/s.
 
\begin{figure}
\centering
\includegraphics[width=\columnwidth]{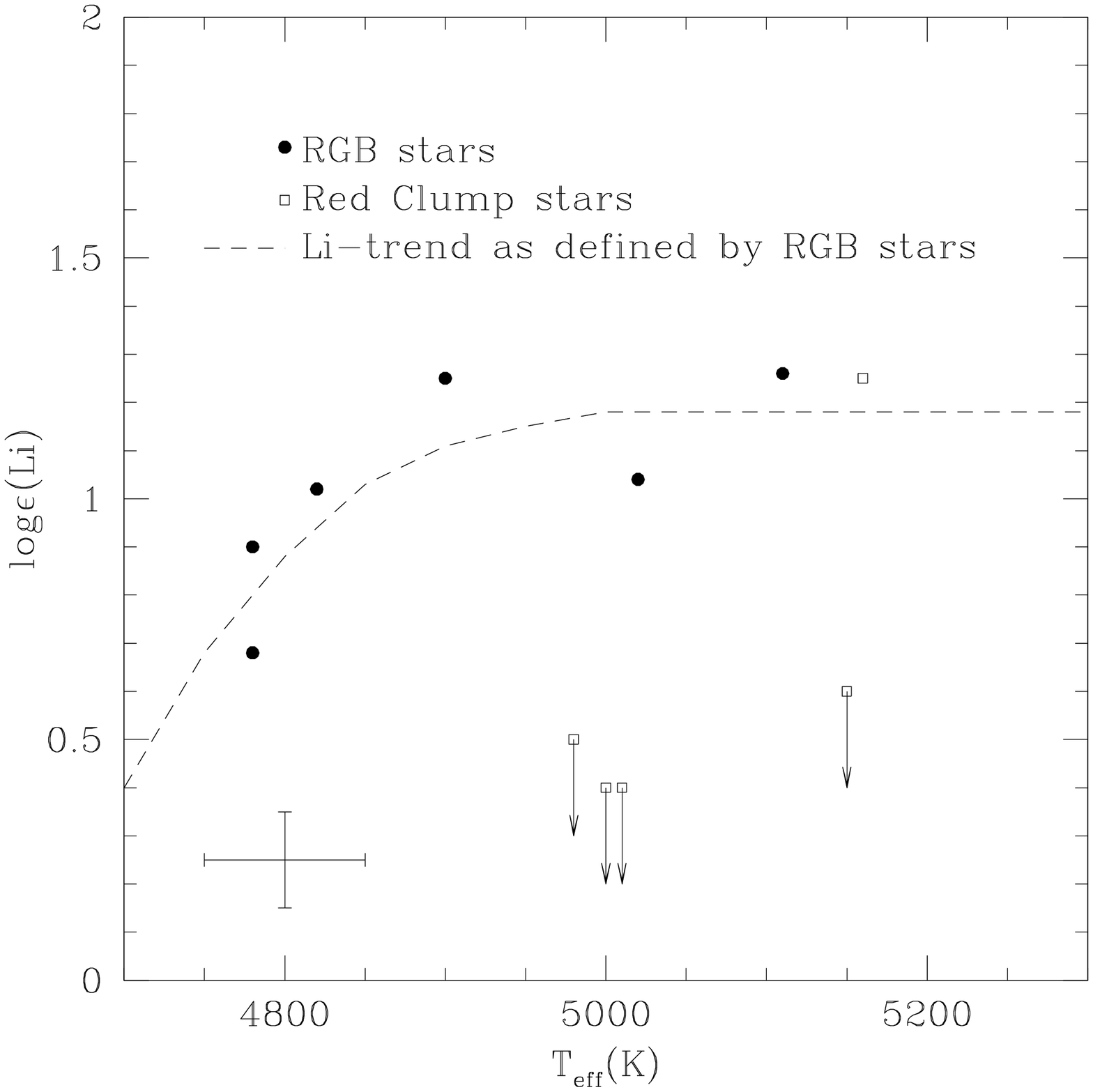}
\caption{Li abundances for observed stars. Evolutionary stage of each
star is indicated. Symbols with arrows represent upper limits.
Typical error bars are indicated.}
\label{f4}
\end{figure}

Once the internal errors associated with the atmospheric parameters were
calculated, we re-derived the abundances of one reference star (\#591),
assumed to represent our sample, by using the following combination of atmospheric parameters:
\begin{itemize}
\item ($\rm {T_{eff}} \pm \Delta (\rm {T_{eff}})$, $\rm {log g}$,  $\rm {v_{t}}$)
\item ($\rm {T_{eff}} $, $\rm {log g} \pm \Delta (\rm {log g})$,  $\rm {v_{t}}$)
\item ($\rm {T_{eff}} $, $\rm {log g}$,  $\rm {v_{t}} \pm \Delta (\rm {v_{t}})$)
\end{itemize}
where $\rm {T_{eff}}$, $\rm {log g}$,  $\rm {v_{t}}$ are the measures
determined in this section.

The difference of abundance between values obtained with the original
and modified values gives
the errors in the chemical abundances due to uncertainties in
each atmospheric parameter. They are listed in Tab.~\ref{t4} (columns 3, 4
and 5). 
Our best estimate of the total error associated to the abundance
measures is calculated as
\begin{center}
$\rm {\sigma_{tot}=\sqrt{\sigma_{EW}^{2}+\Sigma(\sigma_{atm}^{2})}}$
\end{center}
listed in the column 6 of Tab.~\ref{t4}.
Column 7 of Tab.~\ref{t4} is the observed dispersion obtained as described in section 5.
Comparing $\sigma_{\rm obs}$ with $\sigma_{\rm tot}$ we can see
that almost all the elements (Ca, Sc, Fe, Ni, Ba, and Eu) have an abundance that is
homogeneous with no evidence of spread. The two Ba-poor stars (\#1512 and
\#591) and the Li trend will be discussed in the following section.

\begin{table*}
\caption{Evolutionary stage and chemical abundances ($\rm
  {log\epsilon(X)=log(N_{X}/N_{H})+12)}$ of our objects. RGB indicates Red Giant
branch stars, while RC indicates Red Clump stars.}
\label{t3}
\centering
\begin{tabular}{lccccccccc}
\hline
\hline
ID & Ev. Stage & FeI & ${\rm [FeI/H]}$ & LiI & CaI & ScI & NiI & BaII & EuII\\
\hline                      
1512 & RGB & 7.16 & -0.34 &  -      & 6.11 & 2.70 & 6.05 & 1.61 & -    \\
1672 & RGB & 7.12 & -0.38 & 1.02    & 6.19 & 2.73 & 5.82 & 2.01 & 0.45 \\
1827 & RGB & 7.25 & -0.25 & 1.25    & 6.35 & 2.91 & 6.15 & 2.20 & 0.58 \\
1886 & RC  & 7.11 & -0.39 & $<$0.40 & 6.01 & 2.77 & 6.00 & 2.09 & 0.34 \\
2184 & RC  & 7.19 & -0.31 & $<$0.50 & 6.29 & 2.96 & 6.00 & 2.16 & 0.39 \\
 236 & RC  & 7.20 & -0.30 &  -      & 6.26 & 2.75 & 5.88 & 2.11 & 0.35 \\
 238 & RGB & 7.20 & -0.30 & 0.68    & 6.25 & 2.88 & 6.00 & 2.02 & 0.44 \\
2846 & RC  & 7.11 & -0.39 & 1.25    & 6.19 & 3.02 & 5.95 &  -   & 0.82 \\
 299 & RGB & 7.31 & -0.19 & 0.90    & 6.31 & 2.95 & 6.01 & 2.15 & 0.62 \\
3574 & RC  & 7.27 & -0.23 & $<$0.60 & 6.28 & 2.83 & 6.06 & 2.23 & 0.54 \\
3763 & RC  & 7.12 & -0.38 & $<$0.40 & 6.25 & 2.72 & 5.91 & 1.99 & 0.28 \\
3836 & RGB & 7.26 & -0.24 & 1.26    & 6.28 & 2.74 & 5.88 & 2.14 & 0.57 \\
 591 & RGB & 7.13 & -0.37 & 1.04    & 6.16 & 2.65 & 5.82 & 1.61 & 0.34 \\
\hline
Cluster & & & \tiny{[Fe/H]=-0.31} & & \tiny{[Ca/Fe]=+0.16} & \tiny{[Sc/Fe]=-0.01} & \tiny{[Ni/Fe]=+0.00} & \tiny{[Ba/Fe]=+0.07} & \tiny{[Eu/Fe]=+0.23}\\
\hline 
Obs. lines & & 16 & 16 & 1 & 3 & 1 & 2 & 1 & 1 \\      
\hline    
\end{tabular}
\end{table*}

\section{Results of abundance analysis and revised cluster parameters}
\begin{figure}
\centering
\includegraphics[width=\columnwidth]{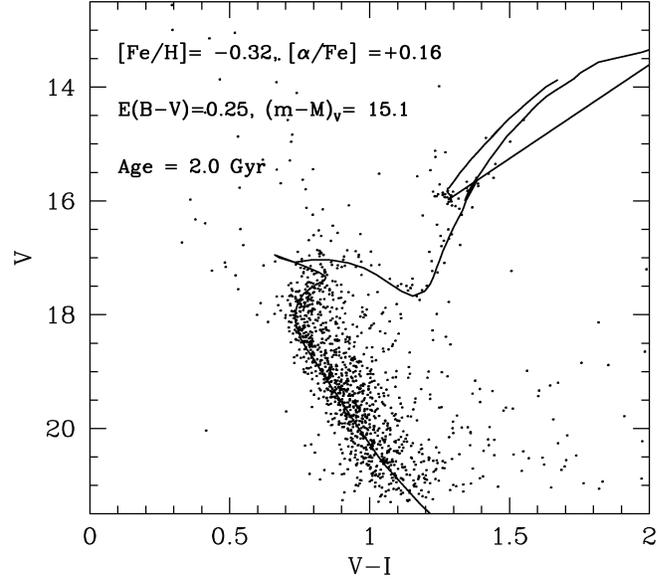}
\caption{Isochrone fitting to the CMD. Derived parameters are indicated.}
\label{f5}
\end{figure}

Chemical abundances we obtained are summarized in Tab.~\ref{t3}
together with the evolutionary stage of each star. This latter is based on 
the star position in the CMD (Fig.~\ref{f5}), where there is a clear division between
RGB and RC stars.
The metallicity of the cluster turns out to be sub-solar with:
\ \\
\begin{center}
${\rm [Fe/H]=-0.31\pm0.02\ dex}$
\end{center}
\ \\
The abundance ratio trends versus [Fe/H] with the exception of Li are shown in
Fig.~\ref{f3}. The bottom right panel report also the [Eu/Ba] content, which
is an indicator of the r-process/s-process element ratio because Eu is an
almost pure r-element, while Ba is mainly produced in s-processes.
Mean abundance ratios were calculated using sigma clipping rejection method.
Rejected measures are plotted as open circles. Mean values are reported in
Fig.~\ref{f3} and Tab.~\ref{t3}.
The cluster turns out to have a solar scaled composition for Sc and Ni, while
Ba is slightly overabundant and Eu is overabundant.
Ca is overabundant and its value (+0.16) is typical of a cluster of this
metallicity. The Ca overabundance found here is consistent with the Ti
overabundance found in F08, as expected from these $\alpha$-elements.
It is interesting to note that two stars (\#1512 and \#591, see Fig.~\ref{f2}) have a very low
Ba content ($\sim$-0.5 with respect to the other stars), that cannot be explained
with measurement errors, which are of the order of 0.1 dex.
The explanation could be an evolutionary effect, but these two stars
are located in the same region of the CMD with respect to the other targets,
so this hypothesis is unlikely. The only remaining reason that occurs to us is
some problem in the data (bad flat-fielding?), but no evidence for this
was found in a further check. In any case \#1512 and \#591 were rejected as suspect
objects as far as Ba content is concerned.\\
The mean [Eu/Ba] value turns out to be +0.17 dex. This is an indication that
Tombaugh 2 could be a intermediate object between the thin and thick disk, according to
\citet[see their Fig.~4]{Mashonkina03}.
In fact in that paper thick disk stars have [Eu/Ba] greater than 0.35 dex,
while thin disk stars have [Eu/Ba]$\sim$0.0 on average. However, its large
distance from the Galactic plane ($\sim$700 pc) make it a sure thick disk member.
According to the measured [Eu/Ba] value, stars of Tombaugh 2 were formed from material more
enriched by r-process elements with respect to the Sun, being the cluster
$\sim$1.5 times richer in Eu that our Star.\\
Li is a very fragile element which is easily destroyed in stellar interiors
at relatively low temperature (2.5$\times$10$^6$ K). During the
life of a star, and in particular during the MS phases,
the Li-rich material that lies near the
surface is circulated downwards where the temperature is high enough for Li
burning to occur. When the star evolves to the red giant phase, the deepening
of the convective envelope brings up to the surface internal matter
which was nuclearly processed and the Li abundance decreases.
Several studies have shown that
further Li depletion occurs after the first dredge-up, 
evidencing the action of an extra-mixing process
(e.g. \citealt{charb95}). This 
can be seen in Fig.~\ref{f4}, where we plot log$\epsilon$(Li) vs. 
temperature. RGB stars are indicated as filled circles, while RC stars are open
squares. One immediately notes that RGB objects have a mean Li abundance
greater than RC stars, for which we could measure only upper limits (see
i.e. stars \#236 and \#591 in Fig.~\ref{f2}).
Only the star \#2846 falls out of this picture because it is appears to be a RC
object but its Li content is high (as high as the other RGB stars). This object could be
explained by error in the photometry, which altered its real position in
the CMD.
Otherwise, we found a clear trend of Li as a function of T$_{\rm eff}$
for RGB stars. Apparently objects hotter than 4900 K (and located in the lower
RGB) have a constant Li abundance (log$\epsilon$(Li)$\sim$1.18). For stars
colder than this limit Li starts to be more and more depleted as the stars 
climb up the RGB (see dashed line in Fig.~\ref{f4}). A very similar behavior
was found by \citet{defreitas} based on UVES spectra of several open
cluster giant members.\\
Having new values for metallicity and $\alpha$-element content, we redetermined
the basic parameters for the cluster using Padova isochrones \citep{marigo08} 
and our V,I photometry.
The apparent distance modulus we found ((m-M)$_{\rm V}$=15.1) is in good
agreement with the one from FeI/II ionization equilibrium.
According to our fit (see Fig.~\ref{f5}), Tombaugh 2 has a reddening of
E(B-V)=0.25 (E(V-I)=0.32) and an age of 2.0$\pm0.1$ Gyr, which is in agreement the value given by F08.

\begin{table*}
\caption{Internal errors associated to the chemical abundances. See text for
more details.
}
\label{t4}
\centering
\begin{tabular}{lcccccc}
\hline
\hline
El. & $\sigma_{\rm EW}$ & $\Delta$T$_{\rm eff}$=50 K & $\Delta$log g=0.06 &\
$\Delta$v$_{\rm t}$=0.02 km/s & $\sigma_{\rm tot}$ & $\sigma_{\rm obs}$\\
\hline                      
$\Delta$${\rm [FeI/H]}$          & 0.03 & 0.06 & 0.02 & 0.02 & 0.07 & 0.07\\
$\Delta$${\rm log\epsilon(Li)}$  & 0.10 & 0.01 & 0.00 & 0.01 & 0.10 & $>$0.2 \\
$\Delta$${\rm [CaI/FeI]}$        & 0.06 & 0.01 & 0.00 & 0.01 & 0.06 & 0.05\\
$\Delta$${\rm [ScI/FeI]}$        & 0.10 & 0.01 & 0.00 & 0.01 & 0.10 & 0.08\\
$\Delta$${\rm [NiI/FeI]}$        & 0.07 & 0.02 & 0.01 & 0.02 & 0.08 & 0.09\\
$\Delta$${\rm [BaII/FeI]}$       & 0.06 & 0.01 & 0.03 & 0.00 & 0.07 & 0.06\\
$\Delta$${\rm [EuII/FeI]}$       & 0.05 & 0.02 & 0.02 & 0.00 & 0.06 & 0.07\\
\hline     
\end{tabular}
\end{table*}

\section{Comparison with previous metallicity studies} 

We first compare our findings with those of \citet{bro96} and \citet{frie02}.
The former conducted a high resolution (R $\sim$ 34,000)
study of To2 with the CTIO 4-m telescope
and found [Fe/H] = -0.40$\pm$0.25 for E(B-V)=0.4 or
[Fe/H]=-0.5$\pm$0.25 for E(B-V)=0.3 based on three stars,
with [Fe/H]=-0.2,-0.4, and -0.6.
The Brown abundance analysis showed that To2 has a reddening of E(B-V)=0.3-0.4. 
The metallicities of their three stars are on the low end of the range of our
derived [Fe/H] values.
However, the more metal-rich of the three Brown et al. stars is ruled out as a 
cluster member based on our radial velocity criterion.
Thus, accounting for this, the Brown et al. metallicity would be lower than
our mean value, by -0.2 to -0.3 dex, but still within 1$\sigma$
of our value.

\citet{frie02} also obtained spectroscopic abundances for To2, 
using the  CTIO 4-m/ARGUS, which yielded a much
lower resolution ($R \sim 1300$) than either our study, F08
or that of \citet{bro96}, and
with metallicities determined from spectral indices. Friel found [Fe/H] =
$-$0.44 $\pm$ 0.09 for To2 from a sample of 12 member stars, and with the
individual measurements ranging from$-$0.28 to $-$0.65. 
Their value is in good agreement with ours.

We now turn to a comparison with F08. Recall that they also used the VLT,
obtaining both high resolution UVES data as well as lower resolution GIRAFFE
data. However, their UVES spectra were of very low S/N, $\sim$15-20, while GIRAFFE 
data were taken in a much more crowded region of the spectrum - 4750-6800\ \AA - than we
employ here - 6500-6800\ \AA. Their GIRAFFE resolution (R=26000) is only slightly
higher than ours.
They derived detailed abundances for 18 cluster members (4 from UVES data),
i.e. with photometry consistent with being RGB or RC stars and with
velocities in the range 121$\pm$4 km/s.
They found evidence for two populations of To2 stars, with a metal-poor group
(mean [Fe/H]=-0.28) and a metal-rich group (mean [Fe/H]=-0.06), with a fairly
small dispersion among each group. We refer to these as the MP and MR 
groups subsequently. The total metallicity range covered a wide margin: 0.00 to -0.43.
Their data also suggested distinct Ti abundances for these
2 groups, with the MP group being $\alpha$-enhanced ([Ti/Fe]=+0.36) and 
the MR group solar ([Ti/Fe]=+0.02). In addition, they found that the MP group
was more centrally concentrated than their MR counterparts. 
They argued that their errors were unable to account for these very surprising
results and suggested that they were indeed real. After discussing a
number of possible scenarios to explain their findings, F08 argued
that the most likely was that the centrally concentrated MP group 
represented the true To2 cluster stars and that the MR group was a spatially overlapping and 
kinematically associated but cold stellar stream. They found this 
scenario to be more feasible than the even more dubious possibility
that To2 possessed a real metallicity spread, which would make it unique among
Milky Way open clusters, but also claimed this as a viable alternative.
Finally, they associated To2 as a likely member of the
GASS/Mon stream.

Our results yield a mean metallicity of -0.31$\pm$0.02 from 13 members, 
with values varying from -0.19 to -0.39, all lying in the range of F08's MP
group. This variation is less than 1/2 that covered by all of F08 stars,
already a strong hint that F08 may have underestimated
their errors. Of equal importance, we find no hint of bimodality in the 
metallicity distribution. Our errors, which we estimate to be of the same order
as those estimated by F08, can completely account for our observed Fe abundance
variation. What about stars in common between the two studies? There are 6, and
the detailed comparison is as follows, where we give (our ID: F08 ID, our 
[Fe/H]: F08 [Fe/H]).
(1672:135, -0.38:-0.07), (1827:140, -0.25:0.00), (238:127, -0.30:-0.34), 
(3574:199, -0.23:-0.20), (3763:182, -0.38:-0.03), and (591:164, -0.37:-0.11).
Of these, all but the latter were both observed with GIRAFFE. Only two of the
stars show good agreement, while the other four consistently have lower
metallicities in our study than found by F08, in particular the lone UVES star.
All four of these stars lie in their MR group but now indeed we find them to 
have metallicities that would have placed them in their MP group. The 
differences are substantially larger than expected given the combined error
estimates. Note that all four UVES stars, all with very low S/N, lie in the MR group.

How do we explain the discrepancy between our results and those of F08? Is it
possible to explain it via differences in the stellar parameters of their 2 
groups? We indeed find that their MP stars are on average about 50K cooler,
have log g some 0.3 smaller and v$_{\rm t}$$\sim$0.05 less than
their MR stars. The combined effects of these differences on the derived metal
abundance, based on their estimates of this value to the input atmospheric
parameters, can only explain about 0.05 dex of the 0.22 dex difference between
the two mean metallicities and is thus insufficient. We feel that the most
likely explanation is that F08 seriously underestimated their errors. In 
particular, their UVES spectra are all of very low quality.
On the other hand the spectral regions of GIRAFFE spectra lie 
substantially blueward of our spectra, in much more crowded regions. 
All this makes EWs more difficult to be measured, and so atmospheric parameters
less accurate that those obtained in this paper.

Thus, we do not agree with the main result of F08, viz. that a real abundance
spread exists among To2 stars and/or stars associated with the purported 
stellar stream. There is now no need for such a stream - all stars are simply
To2 members and all have the same metallicity.
Our stars cover an even larger area than those of F08 (see
Figure \ref{f1}) yet there is no sign of any radial metallicity dependence, 
another result they suggested. 
It appears that F08 results on their MP group are actually applicable to
all To2 members.
A further advantage of our metallicity is that it is in much better
agreement with that expected for a cluster at the Galactocentric distance of
To2, which in fact was used as one of the arguments by F08 for preferring the
MP group as representative of the cluster. As they point out,
stars as metal rich as those in their MR group (nearly solar metallicity)
at R$_{\rm gc}\ge$15 kpc are not consistent with the measured metallicities
of red giants in the outer disc \citep{carn05} as well as outer disc
Cepheid stars \citep{yon06}, which both suggest that the median disc
metallicity at this Galactocentric distance should be [Fe/H] $\sim$ -0.4,
similar to our To2 value.
These studies of outer disc tracers
find no stars as metal rich as the F08 MR group, even for the younger Cepheid populations.
These outer disc stars also show enhanced $\alpha$-element abundances. Indeed, 
if F08  results for their MP stars are correct, then To2 stars are Ti-enhanced
($<$[Ti/Fe]$>$=+0.36), and we also find them to be Ca enhanced 
($<$[Ca/Fe]$>$ = +0.16), as found by \citet{carn05} and \citet{yon06} for outer disc
stars. However, we do not find any trend of [Ca/Fe] vs. [Fe/H], as F08 found for
Ti. Again, evidence for a single chemical composition for To2 stars, as we find,
appears stronger and much more tractable than any of the F08 alternatives.
Finally, we note that the slight Ba overabundance measured by us is in perfect
agreement with the results of \citet{dorazi} who reported
the discovery of a trend of increasing barium abundance with decreasing age 
based on a large sample of Galactic open clusters. Actually, the inclusion
of To2 in the sample allows for better sampling of the age interval between
2 and 4 Gyr.

To conclude, 
we have no additional evidence for or against F08's contention that To2
is indeed a likely member of the GASS/Mon stream. However, their argument that
added weight is given to this possibility by their observed To2 metallicity
spread is now rebuked. They argued that, since NGC 2808 is also believed to be
a member and it shows evidence for an internal population dispersion (Piotto et al.
2007), the fact that To2 also shows such a dispersion favors its GASS/Mon membership.

\section{Discussion and summary}

The existence of multiple populations in Galactic globular clusters and some
Large Magellanic Cloud clusters is now well established \citep{pio08,marino08,marino09,mil09}. 
Indeed, $\omega$ Centauri is now known to possess at least 5 populations
showing different ages and abundances \citep{vil07}. However, certainly not
all such clusters show this phenomenon, at least not in a clear way. Indeed, the dominant characteristic that
seems to be in common among such clusters is their mass - only the most massive
clusters are involved. Given the current paradigm concerning the origin of 
these multiple populations - retention of gas left over and polluted by a first
generation of star formation to form a second, chemically distinct population -
obviously requires sufficient cluster mass to retain the required SN and 
AGB or massive star wind ejecta. The required mass is estimated to be some
10$^5$M$_\odot$ \citep{mie08}.
Thus, only the most massive Galactic globulars fulfill this 
requirement (although note that at least one less massive cluster, M4 at
10$^4$ M$_\odot$,
also shows multiple populations, \citealt{marino08}). The case of the LMC
clusters is more uncertain, but it is clear that the phenomenon is 
correlated with mass. On the other hand, Galactic open clusters generally do not
exceed masses of $10^3 - 10^4M_\odot$ so that the possibility of retaining any first
generation ejecta is very unlikely - any gas that remains after the first 
star formation episode is subsequently quickly removed by the ejecta itself.
Thus, one does not expect to find multiple populations in open clusters.
To2 is a typical open cluster in this respect and certainly does not exceed 
a few $\times 10^3M_\odot$.

The finding by F08 that To2 possessed an intrinsic metallicity spread (or was a superposition
of an outer disc cluster well out of the plane and a cold stellar stream with
exactly the same velocity but distinct chemistry) was then understandably 
met with some incredulity and at the very least required independent 
corroboration. Hence the present study. 

We obtained independent data but using
the same telescope and spectrograph as they did. However, we deliberately 
achieved superior data quality - in both S/N and by observing in a longer
wavelength regime where line crowding was significantly reduced. Our reduction
and analysis procedures were virtually identical to theirs. 

To summarize, we obtained
high resolution VLT+GIRAFFE spectra in the optical 
for a number of stars in the Tombaugh 2 field. Radial velocities and position
in the color-magnitude diagram are used to assess cluster membership. The
spectra, together
with input atmospheric parameters and model atmospheres, are used
to determine detailed chemical abundances for a variety of elements in 13
stars confirmed as members.

We derive a mean metallicity  [Fe/H]$=-0.31 \pm 0.02$, with no evidence
for an intrinsic abundance dispersion. 
We find Ca, Ba, and Eu to be enhanced while Ni and Sc are
solar. Li abundances decrease with T$_{\rm eff}$ on the upper giant branch and
maintain a low level for red clump stars. The mean metallicity we derive is
in good agreement with that expected from the radial abundance gradient in the
disc for a cluster at its Galactocentric distance.
The surprising result found by F08, viz. for 2 distinct abundance
groups within the cluster, implying either a completely unique open cluster
with an intrinsic metallicity spread, or a very unlikely superposition of a
cold stellar stream and a very distant open cluster,
is not supported by our new data. 
We suspect that the F08 data was subject to substantially larger errors than
they estimated, especially given the low S/N of their UVES spectra and their
much bluer wavelength range, which was plagued by line crowding.
To2, instead of being a unique cluster, is found to be
a normal representative of its class.

\begin{acknowledgements}

E.C., S.V. and D.G. gratefully acknowledge support from the Chilean 
{\sl Centro de Astrof\'\i sica} FONDAP No. 15010003.
E.C. and D.G. are also supported by the Chilean Centro de Excelencia en Astrof\'\i sica
y Tecnolog\'\i as Afines (CATA).

\end{acknowledgements}

\bibliographystyle{aa}
\bibliography{ms.bib}

\clearpage

\begin{table*}
\caption{The 22 non-member stars and the 2 members with bad spectra}
\label{t5}
\centering
\begin{tabular}{lcccc}
\hline
\hline
ID1 & $\alpha$($^0$) & $\delta$($^0$) & V & RV$_{\rm H}$\\
\hline                      
1160 & 105.807125 & -20.882889 & 15.735 &  74.5 \\
 153 & 105.771708 & -20.879250 & 16.990 &  61.1 \\
1530 & 105.674958 & -20.847750 & 15.448 &  98.9 \\
1573 & 105.888125 & -20.841778 & 15.781 &  59.8 \\
 169 & 105.814208 & -20.873833 & 15.486 &  90.0 \\
1755 & 105.748833 & -20.830417 & 14.790 & 126.9 \\
2185 & 105.722083 & -20.800333 & 16.286 &  96.3 \\
2343 & 105.712958 & -20.787528 & 15.966 & 104.0 \\
2403 & 105.892917 & -20.781333 & 15.596 &  64.8 \\
 250 & 105.789333 & -20.850444 & 17.132 &  87.2 \\
2510 & 105.861958 & -20.771361 & 17.359 &  18.3 \\
 266 & 105.885417 & -20.846667 & 15.820 & 130.9 \\
2822 & 105.703542 & -20.736306 & 16.412 & 133.5 \\
2827 & 105.836417 & -20.735028 & 17.007 &  38.5 \\
2911 & 105.678292 & -20.725833 & 15.485 & 109.1 \\
2975 & 105.742917 & -20.717972 & 15.789 & 105.6 \\
3332 & 105.751292 & -20.778472 & 16.991 & 130.6 \\
3562 & 105.768583 & -20.821722 & 15.877 &  51.3 \\
 390 & 105.755542 & -20.824139 & 16.032 & 114.9 \\
 472 & 105.853083 & -20.803694 & 15.367 &  53.6 \\
 589 & 105.847417 & -20.780333 & 16.109 & 135.0 \\
 650 & 105.891542 & -20.766750 & 15.736 & 107.2 \\
\hline
4708 & 105.808750 & -20.879833 & 16.899 & 121.8 \\
 651 & 105.675708 & -20.768111 & 15.921 & 121.7 \\
\hline     
\end{tabular}
\end{table*}

\clearpage

\begin{table*}
\caption{The linelist. EWs are reported in m$\AA$}
\label{t6}
\centering
\begin{tabular}{lcrrrrrrrrrrrrr}
\hline
\hline
El.& ${\rm \lambda (\AA)}$&1512& 1672 & 1827 & 1886 & 2184 &  236 &  238 & 2846 &  299 &  3574 &  3763 &  3836 &   591\\
\hline                      
FeI  & 6494.98 & 154.5 & 225.3 &    -  & 188.8 & 203.3 & 180.1 & 224.5 &   -   & 243.9 & 180.9 & 191.7 & 186.0 & 185.3\\
FeI  & 6498.94 &  80.1 &   -   &    -  &   -   &   -   &  91.2 & 109.3 &   -   & 122.7 &  92.6 &   -   &   -   &   -  \\
FeI  & 6533.93 &   -   &  56.9 &  60.3 &  53.8 &  54.2 &   -   &  61.5 &  46.2 &  75.5 &  55.0 &  49.3 &  54.9 &  46.3\\
FeI  & 6569.22 &  88.1 &   -   &  93.2 &   -   &   -   &  92.8 &  92.9 &   -   & 103.4 &  94.8 & 104.3 &  86.8 &  88.9\\
FeI  & 6593.87 &   -   & 122.3 &    -  & 106.0 &   -   & 104.4 & 129.7 &   -   & 127.3 &  95.3 &   -   &   -   & 108.0\\
FeI  & 6608.03 &  26.6 &  47.4 &  51.0 &  39.1 &  41.4 &  32.3 &  54.4 &   -   &  59.3 &   -   &  41.0 &  37.2 &   -  \\
FeI  & 6627.54 &  24.5 &  40.3 &  38.6 &  32.2 &  35.1 &  26.6 &  47.9 &   -   &  44.9 &   -   &  41.7 &  35.2 &  34.0\\
FeI  & 6646.93 &  20.2 &  29.0 &  30.7 &  23.4 &   -   &   -   &  37.2 &   -   &  42.2 &   -   &  28.1 &   -   &  26.5\\
FeI  & 6677.99 & 134.6 & 171.4 & 172.3 & 151.5 & 160.4 & 158.7 & 176.2 & 140.3 & 186.0 & 148.0 & 153.7 & 153.3 & 142.9\\
FeI  & 6713.74 &  -    &  29.5 &    -  &   -   &   -   &   -   &   -   &  26.5 &   -   &   -   &   -   &   -   &   -  \\
FeI  & 6726.66 &  46.5 &  58.3 &  56.2 &  50.2 &  55.3 &  61.6 &  59.9 &  40.0 &  60.8 &  53.5 &  50.4 &  54.8 &  46.4\\
FeI  & 6733.15 &  27.6 &  35.8 &  38.0 &  30.9 &  36.7 &  30.4 &  36.8 &  27.4 &  38.2 &  32.1 &  30.5 &  29.9 &  31.7\\
FeI  & 6739.52 &  25.8 &  50.5 &  49.7 &   -   &  40.6 &   -   &  53.5 &   -   &  54.9 &   -   &  34.9 &  39.8 &  32.8\\
FeI  & 6750.15 &  88.8 & 107.2 & 111.2 & 100.5 & 105.3 & 103.1 & 116.6 &  93.4 & 118.3 &  91.6 &  98.4 &  96.0 & 100.4\\
FeI  & 6752.71 &  42.1 &   -   &  54.3 &  43.2 &  50.6 &  44.4 &   -   &   -   &  61.3 &  42.4 &  41.0 &  45.7 &  40.5\\
FeI  & 6806.84 &  30.5 &   -   &    -  &  53.9 &  60.9 &  60.9 &  67.7 &  39.6 &  77.7 &  59.6 &  52.9 &  56.8 &  59.8\\
FeII & 6456.38 &  70.3 &  63.2 &  67.5 &  73.0 &  79.2 &  64.4 &  57.0 &   -   &  71.9 &  68.5 &  66.9 &   -   &   -  \\
CaI  & 6455.60 &   -   &   -   &   -   &   -   &   -   &  80.9 &  91.9 &   -   &  89.5 &   -   &   -   &   -   &   -  \\
CaI  & 6471.66 & 101.4 & 124.4 & 128.5 &   -   & 110.7 & 112.2 & 126.4 & 110.3 & 135.1 & 117.6 & 114.9 & 109.7 & 107.4\\
CaI  & 6499.65 &  82.1 & 105.1 & 112.8 &  92.2 & 112.4 &  93.6 & 111.4 &  83.5 & 122.4 &  91.5 & 101.9 & 101.5 &  96.3\\
ScII & 6604.60 &  50.4 &  65.9 &  69.1 &  67.6 &  72.4 &  60.9 &  71.3 &  75.1 &  73.7 &  44.5 &  61.9 &  52.6 &  58.6\\
NiI  & 6767.77 & 107.0 & 114.0 & 124.0 & 110.3 & 119.3 & 105.7 &   -   & 109.5 & 127.2 & 111.1 & 106.5 & 102.5 & 109.0\\
NiI  & 6772.31 &  56.2 &  63.2 &  78.4 &  70.3 &  62.4 &  57.2 &  69.4 &  50.2 &  71.0 &  60.3 &  63.6 &  51.8 &  50.9\\
\hline     
\end{tabular}
\end{table*}

\end{document}